# Iterative Time Reversal with Tunable Convergence

**Biniyam Tesfaye Taddese, Thomas M. Antonsen, Edward Ott, Steven M. Anlage**

We propose and test an iterative technique for improving the temporal focusing of a time reversal mirror. A single amplification parameter is introduced to tune the convergence of the iteration. The tunable iterative technique is validated by tests on an experimental electromagnetic time reversal mirror, as well as on a novel numerical model.

*Introduction:* Spatiotemporal focusing of waves has applications in fields such as imaging and communication. Time reversal (TR) mirrors have been used to focus waves in both space and time [1]. An ideal TR mirror consists of a wave source located inside a lossless medium that is completely enclosed by a surface of transceivers which record and absorb the signal initially broadcast by the source. Later, the transceivers rebroadcast a time reversed version of the recorded waves, and because of the TR invariance of the lossless wave equation, the waves focus on the location of the source and reconstruct a time reversed version of the original signal. In practice, TR mirrors have several limitations that result in loss of information about the waves broadcast by the source; these include i) limited coverage by the transceivers, and ii) dissipation during the wave propagation (which breaks TR invariance) [2,3].

The first limitation of TR mirrors can be overcome by the use of a reflecting wave chaotic cavity with partial spatial coverage of the transceivers, along with a long recording time [3]. However, the limitation due to dissipation persists, and leads to increasing loss of information as the recording time increases.

The loss of information during the reconstruction results in temporal and spatial sidelobes of the reconstructed pulse. In previous work, we used the compensating technique of exponential-in-time amplification of the rebroadcasted time-reversed signal to partially undo the adverse effects of dissipation, and to enhance the range of sensors which utilize TR mirrors [2, 4]. However, this technique does not improve the temporal focusing of the reconstructed pulse. On the other hand, Ref. [5] has introduced an iterative TR technique which has been shown to be effective in eliminating the spatiotemporal sidelobes of the reconstructed pulse.

The iterative technique can be useful in applications in which TR mirrors could benefit from enhanced focusing [6]. In this paper, we introduce into the iteration method an amplification parameter to compensate for dissipation. By tuning this parameter, we can substantially improve the accuracy and convergence of the iterative focusing technique. This is demonstrated both experimentally and numerically.

*The Iterative Time Reversal Algorithm with a Convergence Parameter:* The iterative TR method was first introduced using acoustic waves [5]. Consider a regular TR mirror operation that involves broadcasting an original pulse, O, into a cavity with a single port. Denote the scattering parameter of the system by H. We call the response signal received at the port the *sona*, S. From now on, all of these signals are considered in the frequency domain, and hence $S_1 = HO + b_0$; the subscript on S indexes the iteration, and $b_0$ is additive white Gaussian noise (AWGN). For a regular TR (which is the 1$^{st}$ step of the iteration), the *sona*, $S_1$, is time reversed (phase conjugated, as $S_1^*$, in the frequency domain) and broadcast back into the cavity to retrieve the reconstructed pulse at the 1$^{st}$ iteration, $R_1 = HS_1^* + a_1 = HH^*O^* + Hb_0^* + a_1$. Here, $a_1$ is AWGN that is picked up during the recording of $R_1$. Note that in the ideal case $b_0 = a_1 = 0$ (no noise) and $|H|^2 = 1$ (no cavity losses), and $R_1$ is thus equal to $O^*$ (i.e., a time reversed original signal). However, if losses are present $|H|^2$ is frequency dependent and less than unity. The iterative algorithm calculates a new *sona* signal, $S_{n+1}$, by subtracting a correction signal, $C_n$, from the previous *sona*, $S_n$, (i.e. $S_{n+1} = S_n - C_n$); and, the algorithm uses the newly calculated *sona*, $S_{n+1}$, to generate a new reconstructed pulse, $R_{n+1}$, iteratively (i.e. $R_{n+1} = HS_{n+1}^* + a_{n+1}$). The correction signal can be interpreted as the part of the *sona* that resulted in the sidelobes during the reconstruction. The correction signal is obtained by first computing the sidelobes in $R_n$ which are given by $R_n - O^*$. Then, the sidelobes during the $n^{th}$ reconstruction, $R_n-O^*$, are time reversed and broadcast into the system to determine the correction signal, $C_n = H(R_n-O^*)^* k + b_n$. Once again, $b_n$ is AWGN that is picked up while $C_n$ is recorded.

The advantage of introducing the parameter k is revealed by the expression for the $n^{th}$ iterated reconstructed pulse $R_n$ that is derived from the previous equations:

$$R_n = [1-(1-HH^*k)^{n-1}]O^* + HH^*(1-HH^*k)^{n-1}O^*$$
$$+ \sum_{j=0}^{n-1}(1-HH^*k)^j Hb_{n-1-j}^*$$
$$- \sum_{j=0}^{n-2}(1-HH^*k)^j HH^*ka_{n-1-j} + a_n \quad (1)$$

Note that for k=1, Eq.1 reduces to the result in Ref. [5]. The goal of the algorithm is to make $R_n$ approach $O^*$ (the time reversed version of the original pulse) as n increases. The first two terms in Eq. 1 show that the convergence of the iteration can be hastened if k is chosen to make $HH^*k$ as close to 1 as possible over the bandwidth of the pulse, and always less than 2. The optimum k value for the fastest convergence of the iteration is found after an initial reference experiment to measure H. The optimum k is dependent on H and the AWGN in the system.

*The Electromagnetic Experimental Setup:* We experimentally test our method on a 1m$^3$ aluminum box resonant cavity with interior scatterers. The box has two electrical ports that are connected to an oscilloscope, and a microwave source (see Fig. 1). Reciprocity between the two ports simplifies the experiment because the connections to the oscilloscope and the source need not be exchanged. Fig. 1 illustrates how the two steps of the regular TR (i.e. the 1$^{st}$ step of the iterative algorithm) are carried out [7]. Although our derivation above assumed a 1-port situation, we expect [7] that it will also work on this 2-port configuration.

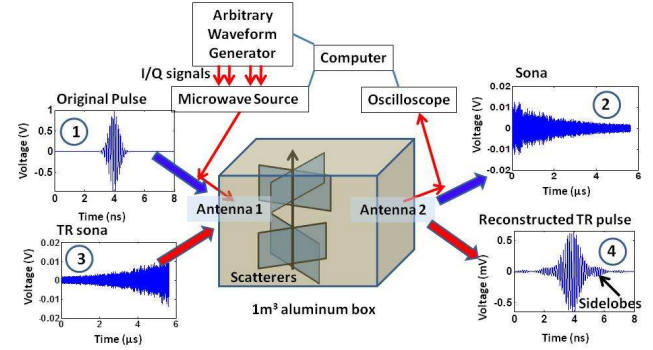

**Fig. 1** *Schematic of the electromagnetic time reversal mirror experiment. During step 1 of the TR mirror, the original pulse is broadcast through antenna 1 (1), and the resulting sona is collected at antenna 2 (2). Next, the time reversed sona is injected into the system at antenna 1 (3) to retrieve the reconstructed time reversed pulse at antenna 2 using spatial reciprocity (4). Experimental data are shown for each step.*

*The Transmission Line Model:* We also numerically test our method by simulation of a model consisting of a driving transmission line that is connected to a number of transmission lines that are connected in parallel with each other (known as a star graph). A sketch of the transmission line model is shown as an inset to Fig. 2. There are 50 transmission lines, each with some specified length and loss constant, terminated by open circuits. The pulse is injected through the driving line. After the pulse reverberates through the lines connected in parallel, it comes back out through the driving line to form the model *sona*. This simple model system captures the essence of multiple pulse trajectories inside complicated 3D scattering systems.

The scattering parameter, H, of this 1-port system (with the port at the driving line) can be analytically determined from the characteristic impedance, the propagation constant, and the terminal reflection coefficient of the lines. Thus, the response to any time domain input signal can be determined by Fourier transforming the input signal, multiplying it by H, and inverse Fourier transforming to get the time domain representation of the output signal.

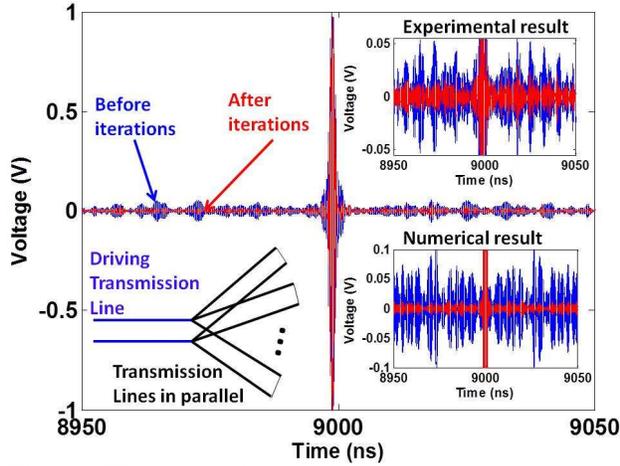

**Fig. 2** *The experimental time reversed pulse reconstructed after 25 iterations using k=110 (red) is overlaid on the pulse reconstructed without the iterative technique (blue); the inset on the top right corner shows a close up view of how the sidelobes are suppressed by the iterative technique experimentally. The inset on the bottom right corner shows a close up view of the suppression of the sidelobes after 25 iterations in the noiseless numerical model using k=2.5; the inset on the bottom left corner shows a schematic of the transmission line model.*

<u>Results:</u> The iterative algorithm is applied to the electromagnetic TR mirror illustrated in Fig. 1. A 1 ns long Gaussian pulse with a center frequency of 7 GHz is used as the original pulse. In Fig. 2, the reconstructed pulse with 1 iteration, and after 25 iterations using k = 110 (empirically determined optimum value of k for fast convergence of $R_n$ to $O^*$) are compared. The inset at the top right corner of Fig. 2 shows a close up view of the resulting sidelobe suppression. The simulation is also carried out using a 1 ns long, 7 GHz input pulse, but without introducing AWGN. The inset at the bottom right corner of Fig. 2 shows the corresponding close up view for the result from the transmission line model after 25 iterations with k = 2.5 (the empirically determined optimum value).

The role of the parameter k in controlling the convergence of the iteration is investigated for different noise levels (standard deviation of the AWGN) in the simulation. Fig. 3 shows the average ratio of the main pulse energy to sidelobe energy versus normalized k for different noise levels; this ratio is a measure of the quality of temporal focusing. The experimental results are compared to the simulations, and show very clear similarity. The optimum k is roughly 2/maximum(HH*), where the maximum is taken over the bandwidth of the original pulse, and the noise; this is also justified by the convergence condition for the first two terms of Eq. 1.

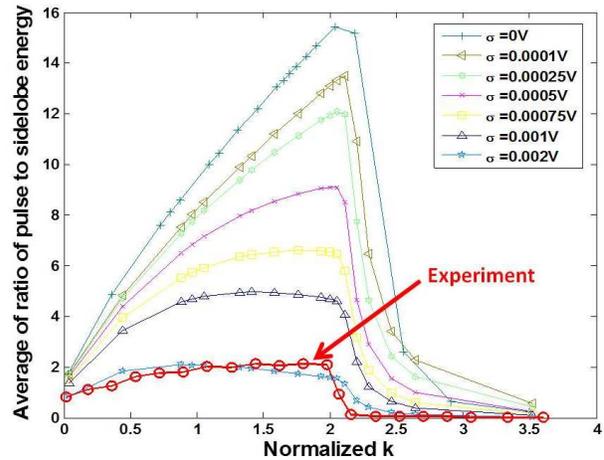

**Fig. 3** *The average ratio of the main pulse energy to sidelobe energy is plotted against normalized k for different noise levels that are labeled by the standard deviation (σ) of the AWGN introduced in the numerical model. The average value is taken over the ratios computed for the first 40 iterations. In addition, similar data from the experiment (shown in Fig. 1) is plotted as red circles. The k value is normalized by multiplying it by the maximum of HH\* (magnitude square of the transfer function) of the system over the bandwidth of the pulse and the noise.*

<u>Conclusion:</u> The iterative time reversal technique is demonstrated experimentally using electromagnetic waves in a microwave wave chaotic cavity, and by simulation. A new amplification parameter is introduced into the iterative algorithm to control the rate of convergence of the iteration. The optimum value of this parameter is dictated by the scattering properties of the system, and, to a lesser extent, by the noise.

**Acknowledgment:**
This work is supported by an ONR MURI Grant No. N000140710734, AFOSR Grant No. FA95501010106, and the Maryland Center for Nanophysics and Advanced Materials.

**Authors' Affiliations:**
B.T. Taddese, T.M. Antonsen, E. Ott and S.M. Anlage (Department of Electrical and Computer Engineering, University of Maryland, College Park, Maryland 20742-3285, USA).
T.M. Antonsen, E. Ott and S.M. Anlage (Department of Physics, University of Maryland, College Park, Maryland 20742-4111, USA).
Email: anlage@umd.edu